\documentclass[12pt,preprint]{aastex}

\newcommand{\degree}{\hbox{$^\circ$}}               
\newcommand{\mum}{$\mu$m}                          
\newcommand{\muJy}{$\mu$Jy}                        

\shorttitle{MIPS 24 and 70\,\mum\ imaging near the SEP field}
\shortauthors{Scott et al.}

\begin{document}


\title{{\it Spitzer} MIPS 24 and 70\,\mum\ Imaging near the South Ecliptic Pole: Maps and Source Catalogs}


\author{Kimberly S. Scott\altaffilmark{1},
        Hans F. Stabenau\altaffilmark{1}, 
        Filiberto G. Braglia\altaffilmark{2}, 
        Colin Borys\altaffilmark{3}, 
        Edward L. Chapin\altaffilmark{2}, 
        Mark J. Devlin\altaffilmark{1}, 
        Gaelen Marsden\altaffilmark{2},
        Douglas Scott\altaffilmark{2}, 
        Matthew D. P. Truch\altaffilmark{1}, 
        Elisabetta Valiante\altaffilmark{2}, and
        Marco P. Viero\altaffilmark{4}}

\altaffiltext{1}{Department of Physics \& Astronomy, University of 
  Pennsylvania, 209 South 33rd Street, Philadelphia, PA 19104, USA}
\altaffiltext{2}{Department of Physics \& Astronomy, University of 
  British Columbia, 6224 Agricultural Road, Vancouver, BC V6T 1Z1, Canada}
\altaffiltext{3}{California Institute of Technology, 1200 East California 
  Boulevard, Pasadena, CA 91125, USA}
\altaffiltext{4}{Department of Astronomy \& Astrophysics, University of 
  Toronto, 50 St. George Street, Toronto, ON M5S 3H4, Canada}


\begin{abstract}
We have imaged an 11.5\,deg$^2$ region of sky towards the South Ecliptic Pole (RA $= 04^\mathrm{h}43^\mathrm{m}$, Dec $= -53\degree40\arcmin$, J2000) at 24 and 70\,\mum\ with MIPS, the Multiband Imaging Photometer for {\it Spitzer}. This region is coincident with a field mapped at longer wavelengths by {\it AKARI} and the Balloon-borne Large Aperture Submillimeter Telescope. We discuss our data reduction and source extraction procedures. The median $1\sigma$ depths of the maps are 47\,\muJy\,beam$^{-1}$ at 24\,\mum\ and 4.3\,mJy\,beam$^{-1}$ at 70\,\mum. At 24\,\mum, we identify 93\,098 point sources with signal-to-noise ratio (SNR) $\ge5$, and an additional 63 resolved galaxies; at 70\,\mum\, we identify 891 point sources with SNR $\ge6$. From simulations, we determine a false detection rate of 1.8\% (1.1\%) for the 24\,\mum\ (70\,\mum) catalog. The 24 and 70\,\mum\ point-source catalogs are 80\% complete at 230\,\muJy\ and 11\,mJy, respectively. These mosaic images and source catalogs will be available to the public through the NASA/IPAC Infrared Science Archive.
\end{abstract}


\keywords{catalogs --- infrared: general --- surveys}


\section{Introduction}
\label{sec:intro}

Understanding the formation and evolution of galaxies is one of the foremost goals of experimental cosmology today. In the redshift range $z\,{\simeq}\,1-3$, massive galaxies go through an evolutionary stage characterized by high rates of star formation, much of which is obscured by dust. Over the past decade, observations at sub-millimeter (sub-mm) and millimeter (mm) wavelengths ($\lambda\sim200-2000$\,\mum) have resulted in the detection of thousands of dust-obscured galaxies at high redshift \citep[e.g.][]{scott02,borys03,greve04,laurent05,coppin06,bertoldi07,greve08,scott08,scott10,perera08,weiss09,dye09,austermann10}. Though these sub-mm/mm galaxies (hereafter SMGs) account for only a small fraction of the cosmic infrared background \citep{puget96,hauser98,fixsen98} at these wavelengths \citep[e.g.][]{wang06,scott08,scott10,devlin09,marsden09,pascale09}, they may contribute significantly to the cosmic star-formation activity at $z\gtrsim2$ \citep{chapman05,aretxaga07,dye08,michalowski10}. While the most luminous sources ($L_{\mathrm{FIR}} \gtrsim 10^{12}\,\mathrm{L}_{\sun}$) are readily detectable over a large range in redshift, owing to a strong negative {\it K}-correction at these wavelengths \citep[e.g.][]{blain02}, the sub-mm/mm data alone provide little insight into the physical properties and redshift distribution of these galaxies, and consequently they need to be identified in other wavebands in order to understand how SMGs fit into the general picture of galaxy evolution.

Over the years, deep complementary multi-wavelength data, particularly at radio and mid-infrared (mid-IR, $\lambda\sim8-50$\,\mum) wavelengths, have proven invaluable for characterizing galaxies detected at sub-mm/mm wavelengths \citep[e.g.][]{pope06,ashby06,hainline09,chapin09,chapin10}. In this paper, we describe 24 and 70\,\mum\ observations taken with the Multiband Imaging Photometer for {\it Spitzer} \citep[MIPS,][]{rieke2004} of a region near the South Ecliptic Pole (SEP), which was recently imaged by the Balloon-borne Large Aperture Submillimeter Telescope \citep[BLAST,][]{pascale08} at 250, 350, and 500\,\mum. This field has one of the lowest cirrus backgrounds at mid-IR wavelengths, with a 24\,\mum\ background of $16\,{\rm MJy}\,{\rm sr}^{-1}$ --- two times lower than that of the COSMOS field and comparable to the Lockman Hole and Chandra Deep Field-South \citep{sanders07}. The BLAST observations have revealed $\sim200$ SMGs in the 8.5\,deg$^2$ field (Valiante et al.~in prep.). The depth of these {\it Spitzer}/MIPS observations ($5\sigma = 250$\,\muJy\,beam$^{-1}$ at 24\,\mum) will allow the identification of mid-IR counterparts for $\sim50\%$ of the BLAST-identified sources out to $z\sim3$. These mid-IR data are also highly complementary to observations at other wavelengths already carried out towards regions within the SEP field, including: mid- and far-IR observations with {\it AKARI} \citep{matsuhara06}; mm-wavelength imaging with AzTEC on the Atacama Submillimeter Telescope Experiment (Hatsukade et al. in prep.), the South Pole Telescope, and the Atacama Cosmology Telescope; and 20-cm observations with the Australia Telescope Compact Array. A 7\,deg$^2$ region within the SEP will also be imaged from 100-500\,\mum\ as part of the {\it Herschel} Multi-tiered Extragalactic Survey (HerMES) Guaranteed Time Key Project.\footnote{http://hermes.sussex.ac.uk}


This paper is organized as follows: In \S\ref{sec:obs} we describe the 24 and 70\,\mum\ observations carried out towards the SEP field. In \S\ref{sec:mosaic} we describe the data reduction process we use to make the 24 and 70\,\mum\ mosaic images. We discuss the source extraction and catalogs in \S\ref{sec:catalogs}, and summarize the final data products in \S\ref{sec:conc}.

\section{Observations}
\label{sec:obs}

The MIPS 24 and 70\,\mum\ observations of the SEP (Program ID 50581) were carried out in a single campaign (MIPS014300) from 2008 September 24-30. The astronomical observational requests (AORs) were designed to be robust against the fast rate of field rotation ($\sim1$\degree\ per day), taking care to provide sufficient overlap to obtain complete sampling at 24 and 70\,\mum. The observations were taken in scan-mode using the medium scan speed (6.5\arcsec\,s$^{-1}$).  We used 160\arcsec\ offsets in the cross-scan direction between forward and reverse scan legs in order to achieve sufficient overlap for the 70\,\mum\ array. Each AOR consisted of nine scan legs with a length of 1.5\degree, and a total of 34 AORs were used to map the field to our target sensitivity ($5\sigma = 250\,$\muJy\,beam$^{-1}$ at 24\,\mum). A total of 88.4\,hrs was spent on these observations.





\section{Mosaic Images}
\label{sec:mosaic}

\subsection{24\,\mum\ Map}
\label{ssec:map24}

We start with the basic calibrated data (BCD; the collection of maps derived from the raw data for each single frame exposure), which are available from the {\it Spitzer} Science Center (SSC) and have been processed using version S18.1.0 of the SSC MIPS 24\,\mum\ pipeline \citep{gordon05, masci05, engelbracht07}. The total number of BCDs from all of the AORs is 66\,093; we exclude 298 frames with unusually high noise --- where the $1\sigma$ root-mean-square (rms) noise is $>10$\,MJy\,sr$^{-1}$ --- and we use the remaining 65\,795 (99.5\%) to make the mosaic. We combine the frames into a single mosaic image using the SSC MOsaicing and Point-source EXtraction ({\sc MOPEX}) software. Before co-adding and combining the BCDs, it is necessary to perform background matching between overlapping frames in order to achieve a common background level. Given the large number of BCDs, we were unable to use the {\sc MOPEX} Overlap pipeline for background matching. Instead, we subtract the mode computed for each frame individually from the original BCDs in order to remove the background prior to mosaicing. Since the background fluctuations for an individual frame are $\lesssim1.5$\% with no strong gradients across the image, the use of higher order differentials is not necessary for background subtraction.

We use the {\sc MOPEX} Mosaic pipeline (version 18.3.3) to interpolate the BCDs onto a common grid, detect and reject outliers, and co-add them into a single image. The frames are first interpolated onto a common grid in RA-Dec (J2000, tangential projection) with $2.45\arcsec$ pixels, using the default interpolation scheme. We then perform multi-frame outlier detection, which identifies and masks both moving objects and cosmic ray strikes. For each pixel in the interpolated grid, the mean and standard deviation of all pixel values from the individual frames are computed, and samples that are $5\sigma$ positive or negative outliers are masked. The frames are then re-interpolated using these masks, and these images are co-added and combined into a single mosaic image.

This initial 24\,\mum\ mosaic image showed noticeable dark latent artifacts oriented in the scan direction over the entire field. Such low-level dark stripes are often seen in 24\,\mum\ scan-mode maps and arise from a 1-2\% reduction in the detector response when the telescope scans over a bright source. With timescales lasting longer than the length of a single AOR, these dark latent artifacts are stable and can be removed by self-calibration. Using the original BCDs, we generate an improved flat-field correction by dividing each frame by the normalized median of all of the BCDs. These represent corrections of $<2.7$\%. The flat-fielded BCDs are then processed in the same way as the original BCDs, resulting in a mosaic image where the dark stripes are largely reduced. These corrections improve the photometry measurements for both point sources and extended sources.

\begin{figure*}
\begin{center}
\includegraphics[width=6.5in]{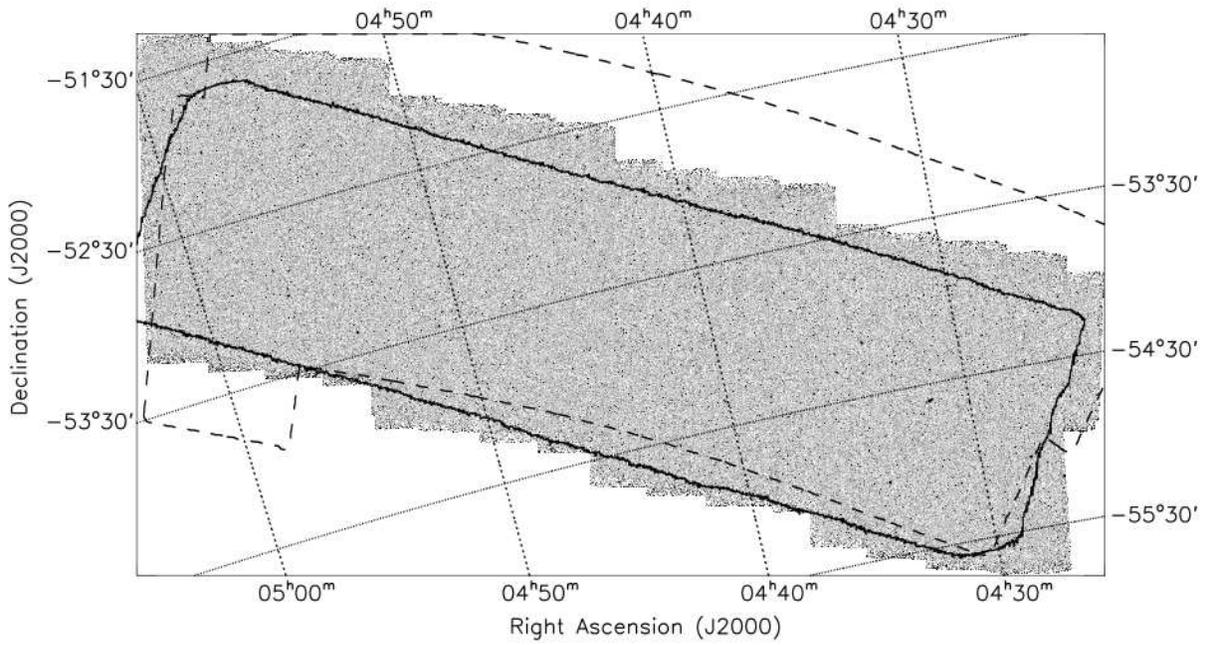}
\caption{The 24\,\mum\ mosaic image of the SEP field. The map is shown on a linear scale ranging from $-0.09$ to 0.3\,MJy\,sr$^{-1}$ (roughly $-3\sigma$ to $10\sigma$). The solid contour shows the overlapping coverage in the 8.5\,deg$^2$ BLAST survey of this field, while the dashed contour indicates the region mapped at 90\,\mum\ by {\it AKARI}.}
\label{fig:map24} 
\end{center}
\end{figure*}

\begin{figure}
\begin{center}
\includegraphics[width=3in]{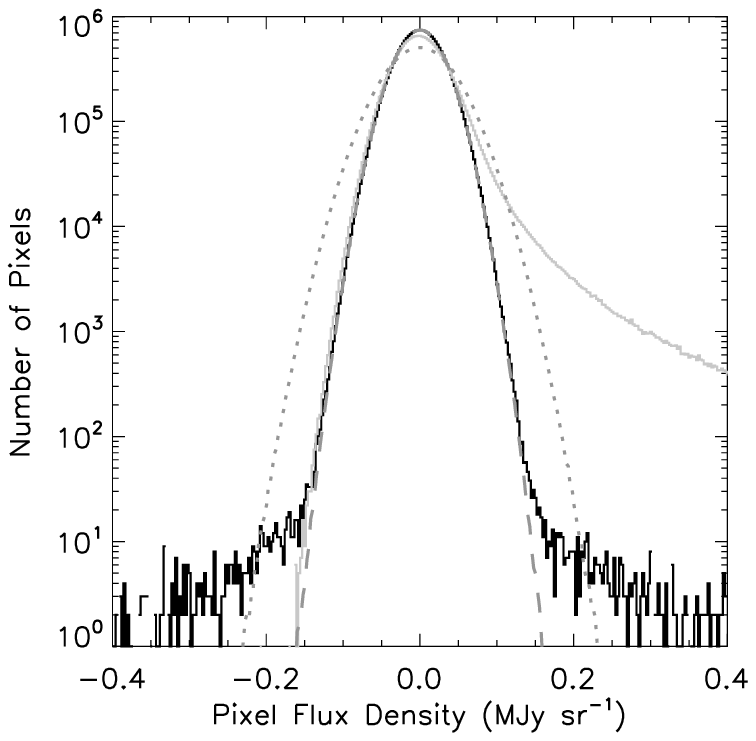}
\caption{Distribution of pixel flux densities in the 24\,\mum\ maps. The solid light-gray histogram shows the flux distribution in the mosaic map. The black histogram shows the flux distribution in the jackknifed noise map. The dotted gray curve is the average flux distribution from simulated noise maps using the original uncertainty values determined from {\sc MOPEX}, and the dashed gray curve shows this distribution after applying a correction factor of 0.68.}
\label{fig:hist24} 
\end{center}
\end{figure}

The 24\,\mum\ mosaic image of the SEP is shown in Figure~\ref{fig:map24}. The map is in units of MJy\,sr$^{-1}$. The {\sc MOPEX} Mosaic pipeline also produces a corresponding uncertainty map (in MJy\,sr$^{-1}$) and a coverage map (number of BCDs averaged for each pixel). However, by studying the pixel flux distribution of the mosaic image (shown in Figure~\ref{fig:hist24} by the solid light-gray histogram), we find that the values in the uncertainty image overestimate the $1\sigma$ noise, as previously noted by other groups \citep[e.g.][]{sanders07}. Since the uncertainty values are used in \S\ref{ssec:cat24} to determine the photometry errors on extracted sources, we apply a correction factor to the uncertainty map produced by the Mosaic pipeline. We construct a realization of the noise in the mosaic map by producing a difference image of overlapping BCDs, alternatively multiplying each successive frame by $\pm1$ before co-adding. The flux distribution for this ``jackknifed'' map is shown as the black histogram in Figure~\ref{fig:hist24}. This technique removes the astronomical signal (both bright and confused sources) from the mosaic image while preserving the properties of the underlying noise. The residual ``noise'' from hot pixels at $\gtrsim |0.15|$\,MJy\,sr$^{-1}$ arises from imperfect subtraction of bright sources. We next generate 20 simulated noise maps from the uncertainty image, assuming that the noise in each pixel is Gaussian distributed with $\sigma$ equal to the pixel value in the uncertainty map. The flux distribution averaged over these noise maps is shown by the gray dotted curve in Figure~\ref{fig:hist24}. We fit the flux distributions of the jackknifed noise realization and the simulated noise maps assuming Gaussian distributions; the ratio of the best-fit $\sigma$ from the jackknifed map flux distribution to that of the average flux distribution from the simulated noise maps is $0.68$. We scale the values in the uncertainty map produced by the Mosaic pipeline by this factor for use in source extraction and all other analyses involving the 24\,\mum~map.

\begin{figure}
\begin{center}
\includegraphics[width=3in]{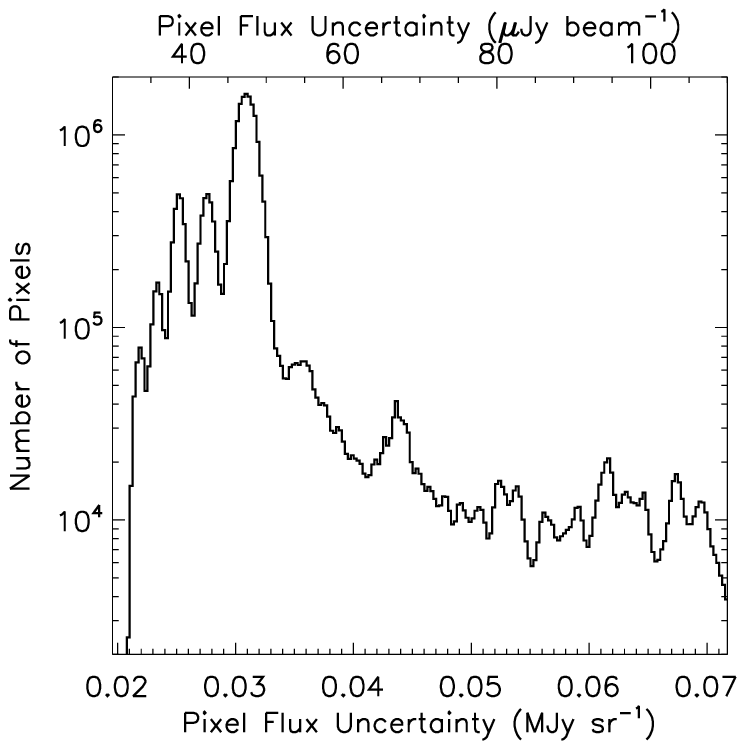}
\caption{Histogram of the pixel uncertainties for the 24\,\mum\ mosaic map (scaled by a factor of 0.68), demonstrating the non-uniform coverage in this map.}
\label{fig:noise24} 
\end{center}
\end{figure}

The total area of the SEP 24\,\mum\ map is 11.8\,deg$^2$, centered at (RA, Dec) $= (04^\mathrm{h}43^\mathrm{m}25.6^\mathrm{s}$, $-53\degree36\arcmin41\arcsec)$. Due to the overlap of the AORs used to map the full region, the coverage in the mosaic image is non-uniform, as demonstrated in Figure~\ref{fig:noise24}. The median $1\sigma$ depth\footnote{We use a conversion factor of 1530\,(\muJy\,beam$^{-1}$)(MJy\,sr$^{-1}$)$^{-1}$, determined by integrating over the 24\,\mum\ point response function (PRF) provided by the SSC.} is 47\,\muJy\,beam$^{-1}$ and ranges from 31-110\,\muJy\,beam$^{-1}$ over the inner 10\,deg$^2$. Assuming a confusion limit (one source per 30 beams) of $\sim200$\,\muJy, estimated from the 24\,\mum\ number counts derived in \citet{papovich04} and \citet{sanders07}, confusion effects on the map properties should be small, but non-negligible.

The spacecraft astrometry is reported to be known to better than 1.4\arcsec. We check for a systematic shift in the astrometry by stacking the 24\,\mum\ map at the positions of 65 stars located within the field (all of which are detected at 24\,\mum).\footnote{From the Smithsonian Astrophysical Observatory (SAO) Star Catalog: http://heasarc.gsfc.nasa.gov/W3Browse/star-catalog/sao.html.}  We find an offset of $(\delta \mathrm{RA},\delta \mathrm{Dec})=(+0.5\arcsec,+0.3\arcsec)$, which given our pixel scale of 2.45\arcsec\ is consistent with zero. The stacked signal is well described by the 24\,\mum\ point response function (PRF) convolved with a Gaussian with $\sigma=1.4$\arcsec. This demonstrates that there are no systematic issues with the astrometry, and the pointing rms errors are as expected.

\subsection{70\,\mum\ Map}
\label{ssec:map70}

For the 70\,\mum\ data, we start with the time-filtered BCD products (fBCDs, total of 66\,098) provided by the SSC. The fBCDs are produced by subtracting the median of the surrounding Data Collection Events (DCEs) as a function of time per pixel, such that the majority of data artifacts caused by variation of the residuals in the slow response and latent artifacts from stimulator flashes are removed. We are left with a total of 63\,168 (95.6\%) fBCDs after excluding those with rms noise $>10$\,MJy\,sr$^{-1}$. As with the 24\,\mum\ data, we remove the background prior to mosaicing by subtracting the mode from each of the frames, and we use the {\sc MOPEX} Mosaic pipeline to combine the frames into a single mosaic image. We interpolate the fBCDs onto a grid with $4.0\arcsec$ pixels (the native pixel scale), and we carry out multi-frame outlier detection as described above for the 24\,\mum\ data, masking samples that are $3\sigma$ outliers (default values in {\sc MOPEX} for 70\,\mum\ data) to produce an initial mosaic image.

Even with the temporal high-pass filter, latent artifacts from stimulator flashes of the internal calibration source, which are correlated by column, are not fully removed; furthermore, the fBCDs provided by the SSC do not preserve calibration for extended sources. To improve the 70\,\mum\ image, we use a median column filter on the data \citep{frayer06a}, starting from the original BCDs and utilizing the Germanium Reprocessing Tools ({\sc GeRT}) available from the SSC. This column filter introduces negative side-lobes near bright sources, so we redo the filtering in two steps: 1) starting with the initial mosaic made from the fBCDs we identify the brightest 10\% of sources in the map using the Astronomical Point-Source Extractor ({\sc APEX}) software; 2) we then use the {\sc GeRT} to column filter the original BCDs with these sources masked. These steps further suppress latent artifacts and improve the calibration for extended sources. After refiltering the BCDs, we perform a background subtraction and use the {\sc MOPEX} Mosaic pipeline to combine them into a single image as described in the previous paragraph.

\begin{figure*}
\begin{center}
\includegraphics[width=6.5in]{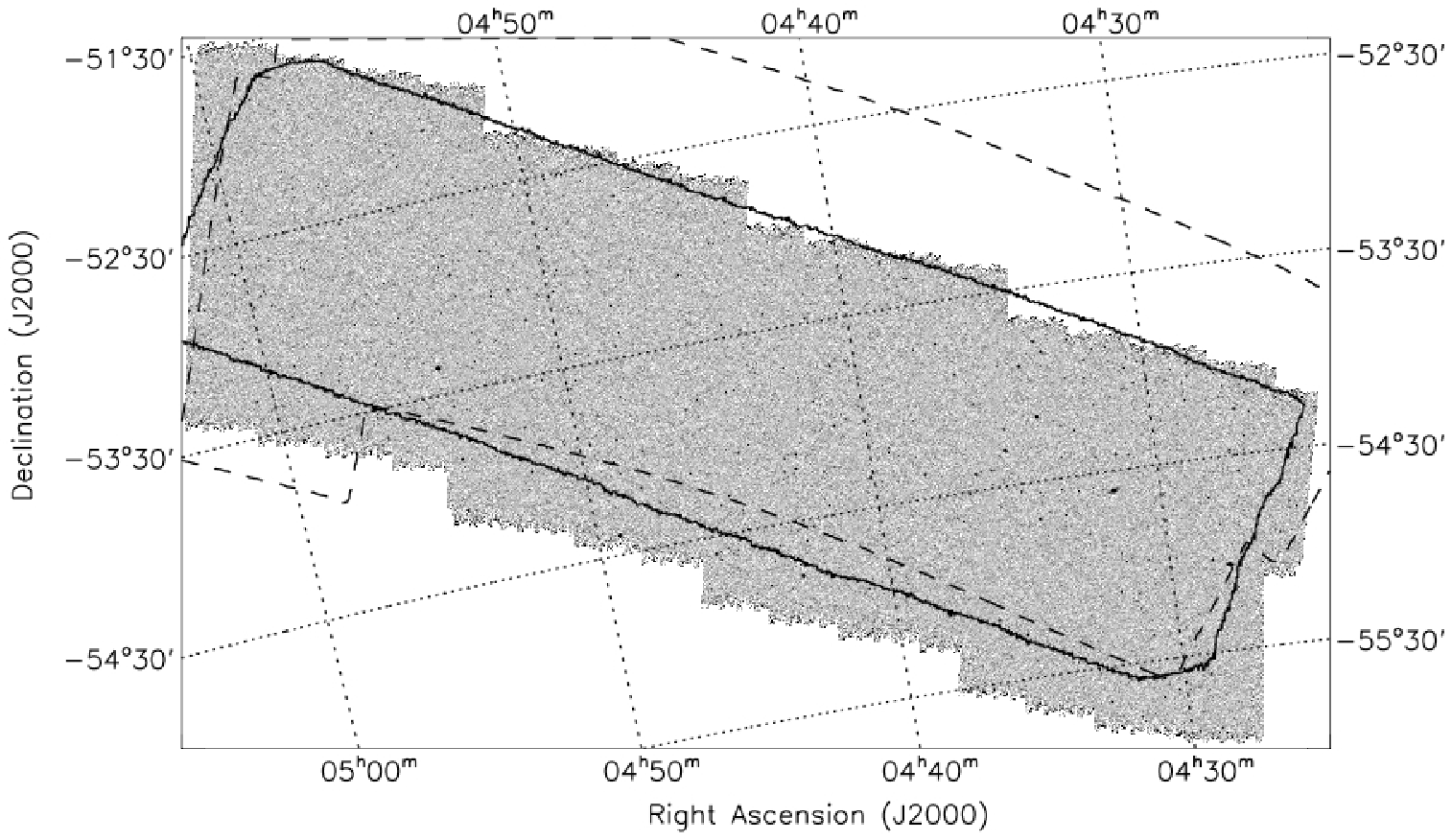}
\caption{The 70\,\mum\ mosaic image of the SEP field. The map is shown on a linear scale ranging from $-0.9$ to 3.0\,MJy\,sr$^{-1}$ (roughly $-3\sigma$ to $10\sigma$). The solid contour shows the overlapping coverage in the 8.5\,deg$^2$ BLAST survey of this field, while the dashed contour indicates the region mapped at 90\,\mum\ by {\it AKARI}.}
\label{fig:map70} 
\end{center}
\end{figure*}

\begin{figure}
\begin{center}
\includegraphics[width=3in]{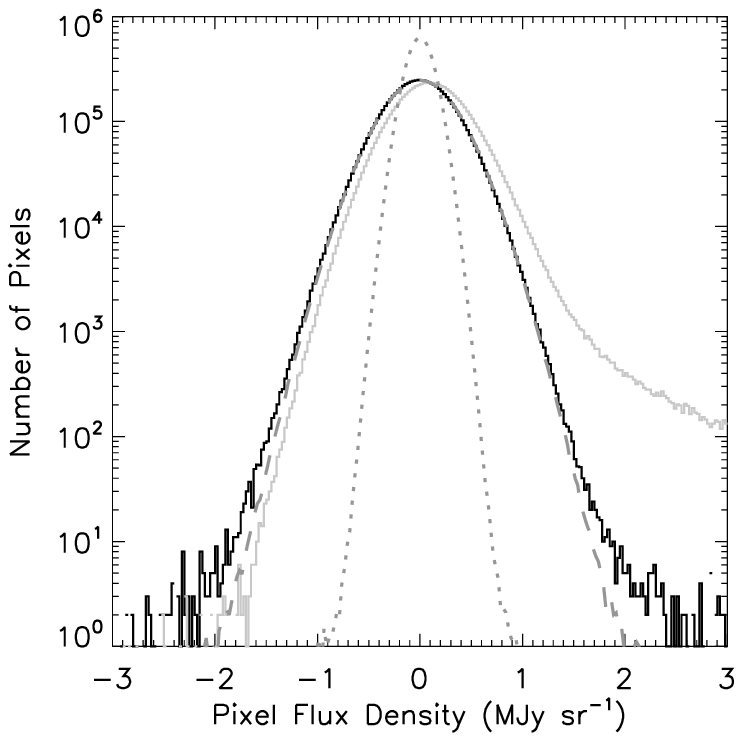}
\caption{Distribution of pixel flux densities in the 70\,\mum\ maps. The solid light-gray histogram shows the flux distribution in the mosaic map. The black histogram shows the flux distribution in the jackknifed noise map. The dotted gray curve is the average flux distribution from simulated noise maps using the original uncertainty values determined from {\sc MOPEX}, and the dashed gray curve shows this distribution after applying a correction factor of 2.55.}
\label{fig:hist70} 
\end{center}
\end{figure}

The 70\,\mum\ mosaic map is shown in Figure~\ref{fig:map70}. As with the 24\,\mum\ mosaic, the corresponding uncertainty image does not provide a good estimate of the $1\sigma$ uncertainty in the map; in this case it significantly underestimates the noise (see Figure~\ref{fig:hist70}). We use the same jackknifing technique as described in \S\ref{ssec:map24} to produce a noise realization for the 70\,\mum\ data, and we determine a correction factor of $2.55$ by comparing the flux distribution of the jackknifed map to that of simulated noise maps made from the original uncertainty image. We use this scaled uncertainty map for all analyses involving the 70\,\mum\ data.

\begin{figure}
\begin{center}
\includegraphics[width=3in]{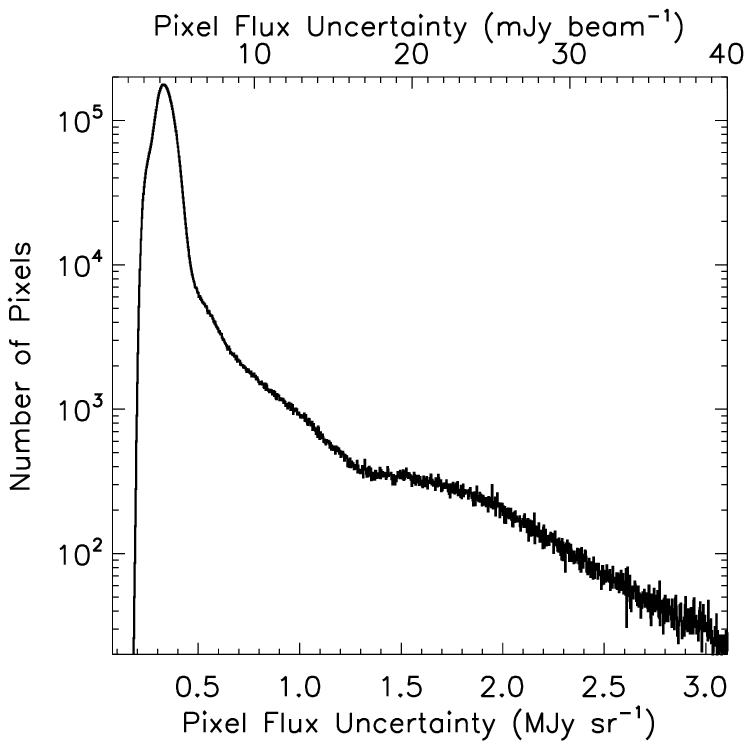}
\caption{Histogram of the pixel uncertainties for the 70\,\mum\ mosaic map (scaled by a factor of 2.55).}
\label{fig:noise70} 
\end{center}
\end{figure}

The total area of the 70\,\mum\ mosaic map of the SEP is 11.5\,deg$^2$, centered at (RA, Dec) $= (04^\mathrm{h}43^\mathrm{m}34.6^\mathrm{s}$, $-53\degree48\arcmin42\arcsec)$. The noise distribution is shown in Figure~\ref{fig:noise70}. The median $1\sigma$ depth\footnote{Using a conversion factor of 12.9\,(mJy\,beam$^{-1}$)(MJy\,sr$^{-1}$)$^{-1}$ determined by integrating over the 70\,\mum\ PRF provided by the SSC.} is 4.3\,mJy\,beam$^{-1}$, ranging from 2.2 to 40\,mJy\,beam$^{-1}$ over the inner 10\,deg$^2$. Given the confusion limit of $\sim8$\,mJy \citep{frayer06a,frayer06b,frayer09}, the effects of confusion on the map properties may be non-negligible.

The spacecraft astrometry for the 70\,\mum\ array is known to better than 1.7\arcsec. Since we have already verified the astrometry in the 24\,\mum\ map, we check for systematics in the 70\,\mum\ map by cross-correlating the 24 and 70\,\mum\ images. We find (as expected) a strong correlation between the two images with an astrometric offset of zero, confirming that the astrometry in the 70\,\mum\ map is good to within the 4\arcsec\ pixel scale.

\section{Source Catalogs}
\label{sec:catalogs}

\subsection{24 $\mu$m Source Extraction}
\label{ssec:cat24}

We use the Astronomical Point-Source Extractor ({\sc APEX}) software within the {\sc MOPEX} package to detect and extract sources from the 24\,\mum\ mosaic and to compute aperture photometry for these sources. We use the point-source probability (PSP) image for source detection and image segmentation. The PSP image is calculated from the background-subtracted mosaic image and the uncertainty image, filtered with the point response function (PRF, Section~\ref{ssec:prf}), and represents the probability at each pixel of having a point source above the noise. Pixels that are $\ge5\sigma$ from the mean are identified and grouped into contiguous pixel clusters; any cluster with $>20$ pixels is run through an iterative process to determine whether to split the pixel cluster into multiple sources. The PRF is then fit to the background-subtracted mosaic image at the source centroids to estimate source fluxes and refine their positions. We allow passive deblending for sources that were split into multiple pixel clusters during image segmentation, where the PRF is simultaneously fit to the blended sources. {\sc APEX} computes two types of uncertainties on the PRF-fitted fluxes. The first represents the naive uncertainty from the fit, which likely underestimates the true flux uncertainty due to correlated errors. The second is computed as the quadrature sum of the data uncertainties within a box the size of the core of the PRF (extending out to $\sim10$\% of the peak). This latter quantity is used to estimate the signal-to-noise ratio (SNR) for the source candidates and generally provides a better estimate of the uncertainty.\footnote{http://ssc.spitzer.caltech.edu/dataanalysistools/tools/mopex/mopexusersguide/91/\#\_Toc253561706}

We select an initial list of source candidates with SNR $\ge5$. For each candidate we consider the PRF fitting to be successful if the $\chi^2$ per degree of freedom (reduced $\chi^2$) is $\chi^2_\mathrm{r}\le3$; this is true for 97\% of the sources. The vast majority of the remaining candidates represent: 1) very bright point sources, many of them known stars in the field; 2) false detections surrounding these bright sources caused by features in the PRF (e.g., the Airy ring); 3) potential bright latent artifacts in the in-scan direction above and below a bright source; 4) extended sources; and 5) false detections arising from extended sources being split into multiple pixel clusters during image segmentation. Since this is a very large field that includes a wide range of sources, it is not possible to select a single group of settings to use for image segmentation that will be optimal in all cases. For this reason we consider the cases above by visually inspecting the mosaic image at the locations of source candidates with $\chi^2_\mathrm{r}>3$, and removing sources that are clearly false detections from the catalog.

\begin{figure}
\begin{center}
\includegraphics[width=3in]{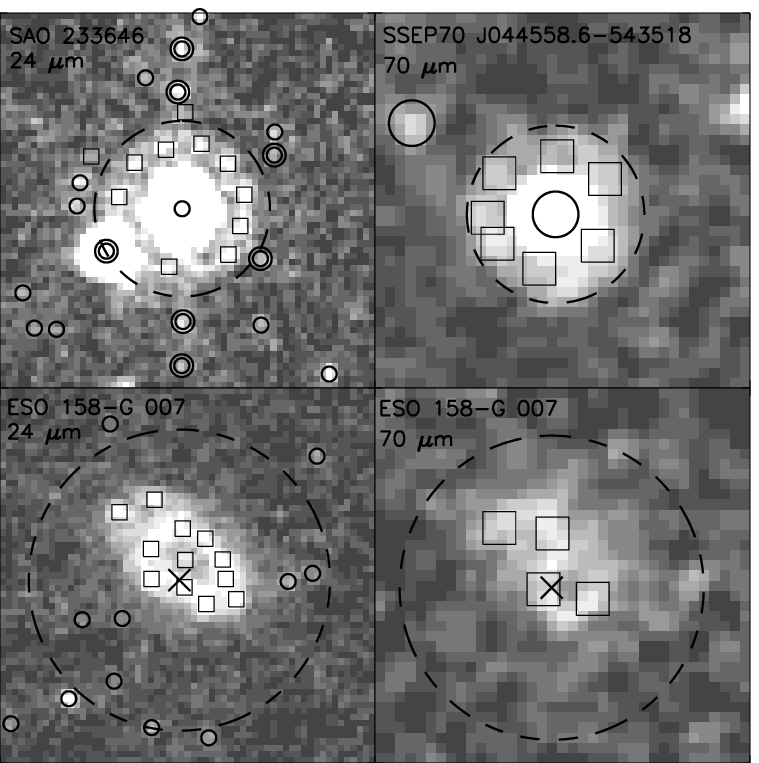}
\caption{Examples of false positives surrounding bright point sources and extended sources in the 24\,\mum\ and 70\,\mum\ maps. All images are $2.5\arcmin\times2.5\arcmin$ and are shown on a linear scale ranging from $-0.03$ to $0.3$\,MJy\,sr$^{-1}$ and $-0.3$ to 3.0\,MJy\,sr$^{-1}$ for the 24\,\mum\ and 70\,\mum\ data, respectively. See the text for a full description of this figure.}
\label{fig:fpexp} 
\end{center}
\end{figure}

Due to the settings used for image segmentation, false detections surrounding bright point sources arise outside of the first Airy ring ($>20$\arcsec\ from the peak emission). From visually inspecting the full mosaic map we identify 90 bright point sources possibly surrounded by such false detections. Of these 65 are known stars. We identify false detections as follows: 1) using the {\sc APEX} Quality Assurance (QA) pipeline, we subtract from the mosaic image a model for the PRF features at $>20$\arcsec\ for each of the 90 bright sources, while retaining the center peak emission inside this radius, creating a residual image; 2) we run the same source detection and extraction algorithm as used on the mosaic image for this residual map, creating a ``residual'' catalog; and 3) source candidates in the original catalog that are not detected in the residual catalog are false detections and are excluded in the final 24\,\mum\ catalog. An example of how we identify false positives surrounding bright point sources is given in the upper left panel of Figure~\ref{fig:fpexp}, which shows a 24\,\mum\ postage stamp image centered on the star SAO 233646. The small circles (diameter = 6\arcsec) and boxes mark the positions of all ``sources'' initially identified using {\sc APEX}, where the latter represent those identified as false positives.

We additionally flag sources that remain in the residual catalog, but may also be false detections given their proximity to a bright source. Examples of these sources --- which we do {\it not} remove from the final catalog --- are represented by double circles in the upper left panel of Figure~\ref{fig:fpexp}. These sources fall into three categories: 1) sources that may represent bright latent artifacts located in the in-scan direction (vertical axis in Figure~\ref{fig:fpexp}); 2) sources that may actually be part of the PRF from the nearby bright source (e.g., radially extended artifacts in the PRF from the telescope secondary mirror support, oriented $\sim60$\degree\ from the scan direction); and 3) sources located within a 35\arcsec\ radius of the bright source (black dashed circle in Figure~\ref{fig:fpexp}, enclosing the second Airy ring). Some of these sources may also be false detections, and most will be poorly fit due to their proximity to a bright source. We describe the identification of false detections around extended sources in \S\ref{ssec:ext}.

The final 24\,\mum\ point-source catalog is available in the electronic version of the {\it Astrophysical Journal Supplement Series}, and the first 15 entries are shown in Table~\ref{tab:pcat24}. There is a total of 93\,098 point sources with SNR $\ge5$, after excluding known false detections. Extended sources are discussed in \S\ref{ssec:ext} and listed separately in Table~\ref{tab:ecat}. The number of sources identified in this field is consistent with that found in other surveys; accounting for the expected number of false positives from noise peaks (\S\ref{ssec:fdr}) and incompleteness (\S\ref{ssec:comp}), the number density of sources with 24\,\mum\ flux density $S_{24} > 300$\,\muJy\ is 0.8\,arcmin$^{-2}$, compared to 0.6-0.9\,arcmin$^{-2}$ observed in other deep {\it Spitzer} surveys \citep{papovich04,sanders07}. 

\clearpage

\begin{deluxetable}{lccccccccl}
\tabletypesize{\scriptsize}
\setlength{\tabcolsep}{0.02in}
\rotate
\tablecaption{SEP 24\,\mum\ Point-Source Catalog\label{tab:pcat24}}
\tablewidth{0pt}
\tablehead{
\colhead{Source Name} & \colhead{RA}            & \colhead{Dec}                                &
\colhead{$S_{\mathrm{PRF}} \pm \sigma_{\mathrm{PRF}}$} & 
\colhead{SNR} & \colhead{$\chi^2_\mathrm{r}$} & 
\colhead{$S_{4.9} \pm \sigma_{4.9}$} &
\colhead{$S_{7.4} \pm \sigma_{7.4}$} &
\colhead{$S_{15} \pm \sigma_{15}$} & \colhead{Comment} \\
\colhead{}            & \colhead{(h m s J2000)} & \colhead{(\degree\ \arcmin\ \arcsec\ J2000)} &
\colhead{(\muJy)}                                      &
\colhead{}    & \colhead{}                     & 
\colhead{(\muJy, uncorrected)}       &
\colhead{(\muJy, uncorrected)}       &
\colhead{(\muJy, uncorrected)}       & \colhead{}        
}
\startdata
SSEP24 J042739.3$-$551438 &  04 27 39.39   & $-$55 14 38.0 & $   325 \pm  25$ &   11   &    2.2  & $   158   \pm 14  $ & $   244 \pm 22$ & $   284 \pm  45$ &         \\
SSEP24 J042835.5$-$540316 &  04 28 35.51   & $-$54 03 16.6 & $   454 \pm  22$ &   16   &    1.0  & $   255   \pm 13  $ & $   357 \pm 20$ & $   421 \pm  41$ &         \\
SSEP24 J042939.9$-$554129 &  04 29 39.91   & $-$55 41 29.4 & $   295 \pm  36$ &    6.6 &    1.0  & $   160   \pm 20  $ & $   227 \pm 32$ & $   658 \pm  66$ & D       \\
SSEP24 J042940.7$-$554133 &  04 29 40.80   & $-$55 41 33.1 & $   272 \pm  37$ &    6.0 &    1.0  & $   155   \pm 21  $ & $   275 \pm 32$ & $   747 \pm  67$ & D       \\
SSEP24 J042941.6$-$554127 &  04 29 41.66   & $-$55 41 27.9 & $   459 \pm  36$ &   10   &    1.0  & $   240   \pm 20  $ & $   280 \pm 32$ & $   411 \pm  66$ & D       \\
SSEP24 J043143.5$-$550749 &  04 31 43.57   & $-$55 07 49.4 & $   258 \pm  20$ &   10   &    0.86 & $   138   \pm 12  $ & $   194 \pm 18$ & $   391 \pm  36$ &         \\
SSEP24 J043410.3$-$552132 &  04 34 10.35   & $-$55 21 32.9 & $   225 \pm  25$ &    7.3 &    1.4  & $   110   \pm 14  $ & $    98 \pm 22$ & $    14 \pm  46$ & P       \\
SSEP24 J043413.3$-$552113 &  04 34 13.37   & $-$55 21 13.3 & $  4358 \pm  26$ &  140   &    8.8  & $  2252   \pm 14  $ & $  2948 \pm 22$ & $  4323 \pm  46$ & S       \\
SSEP24 J043657.3$-$545736 &  04 36 57.34   & $-$54 57 36.1 & $   597 \pm  21$ &   23   &    1.5  & $   320   \pm 12  $ & $   429 \pm 18$ & $   659 \pm  38$ &         \\
SSEP24 J044008.5$-$545205 &  04 40 08.60   & $-$54 52 05.5 & $  2348 \pm  27$ &   72   &    1.8  & $  1230   \pm 15  $ & $  1518 \pm 23$ & $  2120 \pm  48$ & S, D    \\
SSEP24 J044009.5$-$545157 &  04 40 09.56   & $-$54 51 57.2 & $   324 \pm  27$ &    9.8 &    1.8  & $   214   \pm 15  $ & $   345 \pm 24$ & $  1843 \pm  49$ & D, P    \\
SSEP24 J044336.2$-$533418 &  04 43 36.26   & $-$53 34 18.5 & $   143 \pm  23$ &    5.2 &    0.79 & $    51   \pm 13  $ & $    63 \pm 20$ &        ---       &         \\
SSEP24 J044541.2$-$533005 &  04 45 41.23   & $-$53 30 05.4 & $   175 \pm  25$ &    5.8 &    0.75 & $    79   \pm 14  $ & $   103 \pm 21$ & $   195 \pm  44$ &         \\
SSEP24 J044938.7$-$531959 &  04 49 38.75   & $-$53 19 59.8 & $   281 \pm  26$ &    9.0 &    0.62 & $   126   \pm 14  $ & $   140 \pm 22$ & $   199 \pm  46$ &         \\
SSEP24 J045209.8$-$531448 &  04 52 09.86   & $-$53 14 48.6 & $   157 \pm  25$ &    5.2 &    1.2  & $    75   \pm 14  $ & $    92 \pm 22$ &        ---       &         \\
\enddata
\tablecomments{Table~\ref{tab:pcat24} is published in its entirety in the electronic edition of the {\it Astrophysical Journal Supplement Series}. A random sample of 15 entries are shown here for guidance regarding its form and content. The first column gives the source name using the International Astronomical Union (IAU) format. The second and third columns list the RA and Dec for each source. The fourth column gives the PRF-fitted flux density and its formal uncertainty. The fifth column gives the SNR estimate, and the sixth column gives the reduced $\chi^2$ for the fit. Columns 7, 8, and 9 list the (uncorrected) aperture fluxes and uncertainties (\S\ref{ssec:apphot}) using 4.9\arcsec, 7.4\arcsec, and 15\arcsec\ radius apertures, respectively. The last column includes comments on the sources as follows: 1) ``S'' - source is a known star; 2) ``D'' - source was passively deblended, i.e. simultaneously fit along with neighboring sources (listed consecutively in the table, having the same $\chi^2_\mathrm{r}$); and 3) ``P'' - source may actually be part of the PRF feature of a nearby bright source, a bright latent artifact, or be poorly fit due to its proximity to a bright source, as described in \S\ref{ssec:cat24}.}
\end{deluxetable}

\clearpage

\begin{deluxetable}{lcccccl}
\tabletypesize{\scriptsize}
\setlength{\tabcolsep}{0.02in}
\rotate
\tablecaption{SEP Extended Source Catalog\label{tab:ecat}}
\tablewidth{0pt}
\tablehead{
\colhead{Source Name} & 
\colhead{RA}            & \colhead{Dec}                                & 
\colhead{$S_{24} \pm \sigma_{24}$} & 
\colhead{$S_{70} \pm \sigma_{70}$} & \colhead{Aperture Radius} &
\colhead{Note} \\
\colhead{}            & 
\colhead{(h m s J2000)} & \colhead{(\degree\ \arcmin\ \arcsec\ J2000)} &
\colhead{(mJy)}                    &
\colhead{(mJy)}                    & \colhead{(\arcsec)}       &
\colhead{}       
}
\startdata
2MASX J04362281$-$5510342        &  04 36 22.76   & $-$55 10 34.4 & $ 11.26  \pm 0.10 $ & $ 112.4  \pm  1.8 $                  &  30 & SSEP70 J043622.7$-$551035 \\
2MASX J04430361$-$5446543        &  04 43 03.56   & $-$54 46 54.2 & $ 28.03  \pm 0.15 $ & $1068.0  \pm  6.9 $                  &  25 & SSEP70 J044303.5$-$544652 \\
2MASX J04453204$-$5434252        &  04 45 32.03   & $-$54 34 25.2 & $  5.43  \pm 0.11 $ & $  68.4  \pm  1.7 $                  &  30 & SSEP70 J044531.9$-$543425 \\
NGC 1602                         &  04 27 54.97   & $-$55 03 27.8 & $ 48.27  \pm 0.28 $ & $ 428.2  \pm  7.0 ^\star$             &  90 &                         \\
2MASX J04314165$-$5455393        &  04 31 41.65   & $-$54 55 39.3 & $  3.68  \pm 0.06 $ & $  36.9  \pm  1.2 $                  &  20 & SSEP70 J043141.3$-$545540 \\
NGC 1596                         &  04 27 38.11   & $-$55 01 40.1 & $ 11.83  \pm 0.38 $ & $ <            24 $                  & 120 &                         \\
2MASX J04425888$-$5432544        &  04 42 58.86   & $-$54 32 54.3 & $  4.59  \pm 0.06 $ & $  34.9  \pm  1.6 $                  &  20 & SSEP70 J044258.7$-$543257 \\
2MASX J04451295$-$5427073        &  04 45 12.92   & $-$54 27 06.8 & $  7.78  \pm 0.07 $ & $  84.4  \pm  1.6 $                  &  20 & SSEP70 J044513.0$-$542706 \\
2MASX J04452872$-$5420472        &  04 45 28.72   & $-$54 20 47.4 & $  6.83  \pm 0.06 $ & $ 107.4  \pm  1.4 $                  &  20 & SSEP70 J044528.7$-$542047 \\
2MASX J04342317$-$5441331        &  04 34 23.19   & $-$54 41 33.0 & $ 19.23  \pm 0.07 $ & $ 420.1  \pm  3.6 $                  &  20 & SSEP70 J043423.1$-$544132 \\
2MASX J04480892$-$5410540        &  04 48 08.91   & $-$54 10 53.7 & $  9.01  \pm 0.07 $ & $ 136.0  \pm  2.2 $                  &  20 & SSEP70 J044808.8$-$541054 \\
2MASX J04354249$-$5435532        &  04 35 42.49   & $-$54 35 53.0 & $  5.01  \pm 0.05 $ & $  41.2  \pm  1.1 $                  &  20 & SSEP70 J043542.6$-$543551 \\
NGC 1617                         &  04 31 39.53   & $-$54 36 08.2 & $ 87.23  \pm 0.41 $ & $1107.0  \pm  7.7 ^\star$             & 135 &                         \\
SUMSS J043005$-$543910           &  04 30 05.53   & $-$54 39 10.7 & $  2.08  \pm 0.07 $ & $ <            22 $                  &  20 &                         \\
2MASX J04284373$-$5438274        &  04 28 43.74   & $-$54 38 27.8 & $  5.23  \pm 0.08 $ & $  44.5  \pm  1.4 $                  &  30 & SSEP70 J042843.9$-$543825 \\
ESO 158$-$G 007                  &  04 49 37.27   & $-$53 54 42.5 & $ 18.59  \pm 0.17 $ & $ 165.2  \pm  3.4 ^\star$             &  60 &                         \\
ESO 158$-$G 006                  &  04 48 40.36   & $-$53 54 43.8 & $ 15.18  \pm 0.10 $ & $ 229.9  \pm  2.2 $                  &  30 & SSEP70 J044840.3$-$535443 \\
2MASX J04514200$-$5345126        &  04 51 42.01   & $-$53 45 12.5 & $  8.91  \pm 0.05 $ & $ 126.8  \pm  1.5 $                  &  20 & SSEP70 J045142.1$-$534512 \\
ESO 157$-$G 047                  &  04 39 19.13   & $-$54 12 41.4 & $  4.69  \pm 0.15 $ & $  63.2  \pm  1.6 $                  &  45 & SSEP70 J043919.4$-$541238 \\
ESO 157$-$G 043                  &  04 35 15.47   & $-$54 18 57.2 & $ 58.45  \pm 0.16 $ & $ 720.4  \pm  5.2 ^\star$             &  60 &                         \\
IC 2085                          &  04 31 24.24   & $-$54 25 00.6 & $ 26.68  \pm 0.23 $ & $ 376.2  \pm  4.5 ^\star$             &  75 &                         \\
APMUKS(BJ) B043243.73$-$542450.9 &  04 33 50.84   & $-$54 18 40.5 & $  4.44  \pm 0.10 $ & $  47.9  \pm  1.5 $                  &  30 & SSEP70 J043350.8$-$541838 \\
2MASX J04444398$-$5355395        &  04 44 43.97   & $-$53 55 39.4 & $  3.06  \pm 0.05 $ & $  27.0  \pm  1.1 $                  &  20 & SSEP70 J044443.8$-$535539 \\
2MASX J04410494$-$5402486        &  04 41 04.93   & $-$54 02 48.6 & $  8.51  \pm 0.08 $ & $  86.3  \pm  1.4 $                  &  25 & SSEP70 J044104.8$-$540248 \\
ESO 157$-$G 042                  &  04 35 12.03   & $-$54 12 20.5 & $ 11.15  \pm 0.15 $ & $ 127.6  \pm  3.2 ^\star$             &  60 &                         \\
APMBGC 157$-$064$-$039           &  04 33 13.15   & $-$54 13 57.5 & $  4.10  \pm 0.07 $ & $  42.2  \pm  1.2 $                  &  25 & SSEP70 J043312.9$-$541400 \\
ESO 158$-$G 001                  &  04 41 38.90   & $-$53 54 21.4 & $  8.81  \pm 0.10 $ & $  89.0  \pm  1.8 $                  &  30 & SSEP70 J044138.9$-$535421 \\
NGC 1705                         &  04 54 13.50   & $-$53 21 39.8 & $ 48.09  \pm 0.24 $ & $1175.0  \pm  6.0 ^\star$             &  75 &                         \\
2MASX J04594242$-$5302365        &  04 59 42.41   & $-$53 02 36.5 & $  0.69  \pm 0.06 $ & $ <            20 $                  &  20 &                         \\
ESO 157$-$G 030                  &  04 27 32.60   & $-$54 11 48.1 & $  7.69  \pm 0.15 $ & $ 141.5  \pm  2.0 $                  &  45 & SSEP70 J042732.6$-$541148 \\
2MFGC 03850                      &  04 41 52.62   & $-$53 42 12.1 & $  6.04  \pm 0.07 $ & $  80.2  \pm  1.7 $                  &  25 & SSEP70 J044152.6$-$534211 \\
2MASX J04440985$-$5336563        &  04 44 09.83   & $-$53 36 56.5 & $  3.01  \pm 0.05 $ & $  30.19 \pm  0.99$                  &  20 & SSEP70 J044409.8$-$533653 \\
FGCE 0439                        &  04 48 02.76   & $-$53 26 16.4 & $  9.85  \pm 0.08 $ & $ 112.1  \pm  1.8 $                  &  25 & SSEP70 J044802.6$-$532615 \\
2MASX J04342117$-$5353522        &  04 34 21.20   & $-$53 53 52.4 & $ 24.75  \pm 0.08 $ & $ 266.9  \pm  3.0 $                  &  25 & SSEP70 J043421.3$-$535352 \\
2MASX J04263602$-$5406282        &  04 26 36.04   & $-$54 06 28.2 & $ 14.02  \pm 0.07 $ & $ 165.1  \pm  2.2 $                  &  20 & SSEP70 J042636.1$-$540627 \\
IC 2083                          &  04 30 44.27   & $-$53 58 51.0 & $ 11.68  \pm 0.09 $ & $ 138.8  \pm  1.9 $                  &  25 & SSEP70 J043044.0$-$535850 \\
2MASX J04290665$-$5401202        &  04 29 06.67   & $-$54 01 20.4 & $  8.15  \pm 0.07 $ & $  74.2  \pm  1.6 $                  &  20 & SSEP70 J042906.8$-$540120 \\
ESO 158$-$G 014                  &  04 54 45.75   & $-$53 05 57.5 & $  9.39  \pm 0.11 $ & $ 150.8  \pm  2.2 $                  &  35 & SSEP70 J045445.7$-$530557 \\
2MASX J04283256$-$5359474        &  04 28 32.55   & $-$53 59 47.5 & $ 13.03  \pm 0.06 $ & $ 237.5  \pm  3.1 $                  &  20 & SSEP70 J042832.4$-$535947 \\
2MASX J04505562$-$5312459        &  04 50 55.61   & $-$53 12 45.6 & $ 20.40  \pm 0.13 $ & $ 320.0  \pm  2.5 ^\star$             &  50 &                         \\
2MASX J04334493$-$5346467        &  04 33 44.92   & $-$53 46 46.8 & $ 73.17  \pm 0.07 $ & $ 487.3  \pm  3.5 $                  &  20 & SSEP70 J043344.9$-$534646 \\
2MASX J04293931$-$5352464        &  04 29 39.35   & $-$53 52 46.6 & $ 54.27  \pm 0.07 $ &                                      &  20 &                         \\
ESO 158$-$G 012                  &  04 53 42.79   & $-$52 58 53.6 & $  3.89  \pm 0.10 $ & $  30.5  \pm  1.5 $                  &  30 & SSEP70 J045342.8$-$525852 \\
2MASX J04305049$-$5347492        &  04 30 50.50   & $-$53 47 48.8 & $  5.52  \pm 0.07 $ & $ <           163 $                  &  20 &                         \\
ESO 157$-$G 036                  &  04 29 49.59   & $-$53 48 52.8 & $  1.07  \pm 0.11 $ &                                      &  35 &                         \\
2MFGC 04056                      &  04 57 21.43   & $-$52 46 59.1 & $  3.04  \pm 0.10 $ & $ <            29 $                  &  30 &                         \\
2MASX J04530951$-$5254202        &  04 53 09.53   & $-$52 54 20.4 & $ 17.82  \pm 0.09 $ & $ 321.4  \pm  2.7 $                  &  25 & SSEP70 J045309.4$-$525420 \\
APMBGC 157$-$032$-$065           &  04 29 03.65   & $-$53 44 51.4 & $  2.18  \pm 0.12 $ &                                      &  25 &                         \\
2MASX J05003544$-$5232576        &  05 00 35.42   & $-$52 32 57.4 & $  2.52  \pm 0.08 $ & $ <            22 $                  &  25 &                         \\
IC 2079                          &  04 28 30.82   & $-$53 44 16.5 & $ 28.76  \pm 0.30 $ &                                      &  45 &                         \\
ESO 158$-$G 008                  &  04 49 51.13   & $-$52 59 37.4 & $  3.36  \pm 0.13 $ & $ <            26 $                  &  40 &                         \\
2MASX J04452961$-$5308249        &  04 45 29.63   & $-$53 08 24.8 & $  7.42  \pm 0.06 $ & $  99.9  \pm  1.9 $                  &  20 & SSEP70 J044529.8$-$530822 \\
2MASX J04574760$-$5233553        &  04 57 47.60   & $-$52 33 55.4 & $  4.86  \pm 0.07 $ & $  58.3  \pm  1.6 $                  &  20 & SSEP70 J045747.4$-$523354 \\
2MASX J04540432$-$5242323        &  04 54 04.32   & $-$52 42 32.5 & $  4.59  \pm 0.06 $ & $  49.8  \pm  1.4 $                  &  20 & SSEP70 J045404.1$-$524234 \\
AM 0445$-$525                    &  04 46 12.27   & $-$52 54 48.7 & $  4.16  \pm 0.07 $ & $ <            23 $                  &  20 &                         \\
ESO 157$-$IG 051                 &  04 41 25.15   & $-$52 59 50.4 & $ 26.24  \pm 0.08 $ & $ <           368 $                  &  25 &                         \\
ESO 157$-$IG 048                 &  04 39 25.19   & $-$53 02 57.8 & $  8.30  \pm 0.07 $ &                                      &  20 &                         \\
APMUKS(BJ) B045842.18$-$521729.7 &  04 59 52.60   & $-$52 13 07.1 & $  3.28  \pm 0.07 $ & $  38.2  \pm  1.4 $                  &  20 & SSEP70 J045952.5$-$521303 \\
ESO 157$-$G 049                  &  04 39 36.88   & $-$53 00 45.5 & $166.90  \pm 0.17 $ &                                      &  50 &                         \\
2MASX J04485406$-$5230438        &  04 48 54.07   & $-$52 30 43.5 & $ 13.43  \pm 0.07 $ & $ 201.0  \pm  2.0 $                  &  20 & SSEP70 J044853.9$-$523044 \\
FGCE 0448                        &  04 54 09.47   & $-$52 11 00.7 & $  3.92  \pm 0.09 $ & $  55.0  \pm  1.5 $                  &  25 & SSEP70 J045409.6$-$521059 \\
2MASX J04580461$-$5125420        &  04 58 04.62   & $-$51 25 42.1 & $  5.36  \pm 0.07 $ &                                      &  20 &                         \\
ESO 203$-$G 012                  &  04 57 26.03   & $-$51 22 49.1 & $ 13.53  \pm 0.18 $ &                                      &  35 &                         \\
\enddata
\tablecomments{The extended source catalog. Column 1 gives the source name from the NASA Extragalactic Database (NED). The RA and Dec are listed in columns 2 and 3. The 24 and 70\,\mum\ fluxes and their uncertainties are given in columns 4 and 5, respectively. The 24\,\mum\ fluxes were measured using aperture photometry, and the aperture radius used is listed in column 6. The last column lists the 70\,\mum\ counterpart from the point-source catalog (Table\,\ref{tab:pcat70}), where available. For sources with a 70\,\mum\ counterpart noted in column 7, column 5 gives the PRF-fitted fluxes and uncertainties from the point-source catalog (Table~\ref{tab:pcat70}). For entries marked with a star, the 70\,\mum\ flux was measured using aperture photometry with the aperture radius listed in column 6. Upper limits (5$\sigma$) are listed for sources not in the 70\,\mum\ point-source catalog. For sources with no 70\,\mum\ flux listed, the source lies outside of the coverage region of that map.}
\end{deluxetable}

\clearpage

\subsection{70 $\mu$m Source Extraction}
\label{ssec:cat70}

We use the {\sc APEX} software to detect and extract sources from the 70\,\mum\ mosaic. Unlike the 24\,\mum\ data, we do not use the PSP image for source detection; we find that this smooths the data too much and results in a large number of false detections. Instead, image segmentation is performed on the background-subtracted image. Pixels that are $\ge5\sigma$ from the mean are grouped into contiguous pixel clusters, and clusters with $>70$ pixels are run through the iterative process to determine whether to divide them into multiple sources. As with the 24\,\mum\ sources, the background-subtracted 70\,\mum\ mosaic image is fit to the PRF at the source centroids to measure the source fluxes and positions.

All of the extracted sources have SNR $\ge6$ (estimated from the data uncertainties as described in \S\ref{ssec:cat24}), however, we again note that this value cannot be directly interpreted in terms of statistical significance. We consider the PRF fitting to be successful if $\chi^2_\mathrm{r}\le3$, which is true for 93\% of the sources. The remaining source candidates are primarily bright point sources surrounded by false detections arising from features in the PRF, and extended sources, which are sometimes split into multiple sources during image segmentation. The first case is demonstrated in the upper right panel of Figure~\ref{fig:fpexp}, which shows the 70\,\mum\ postage stamp image centered on SSEP70 J044558.6-543518. Sources initially identified by {\sc APEX} are indicated by the smaller circles (diameter = 18\arcsec) and boxes, where the latter represent false positives surrounding the bright point source and are located within a 35\arcsec\ radius containing the first Airy ring. We visually inspect the full mosaic image and remove any sources that are clearly false detections from the 70\,\mum\ catalog.

The final 70\,\mum\ point-source catalog is available in the electronic version of the {\it Astrophysical Journal Supplement Series}, and the first 15 entries are shown in Table~\ref{tab:pcat70}. There is a total of 891 point sources with SNR $\ge6$, after excluding known false detections. Extended sources are listed separately in Table~\ref{tab:ecat} and discussed in \S\ref{ssec:ext}. The number density of 70\,\mum\ sources with flux density $S_{70}>19$\,mJy observed in the SEP field (80\,deg$^{-2}$) is marginally consistent with that observed in \citet[][60\,deg$^{-2}$]{frayer09} and other 70\,\mum\ surveys \citep{frayer06a,frayer06b}. 

\clearpage

\begin{deluxetable}{lccccccccl}
\tabletypesize{\scriptsize}
\setlength{\tabcolsep}{0.02in}
\rotate
\tablecaption{SEP 70\,\mum\ Point-Source Catalog\label{tab:pcat70}}
\tablewidth{0pt}
\tablehead{
\colhead{Source Name} & \colhead{RA}            & \colhead{Dec}                                &
\colhead{$S_{\mathrm{PRF}} \pm \sigma_{\mathrm{PRF}}$} & 
\colhead{SNR} & \colhead{$\chi^2_\mathrm{r}$} & 
\colhead{$S_{16} \pm \sigma_{16}$} &
\colhead{$S_{28} \pm \sigma_{28}$} &
\colhead{$S_{36} \pm \sigma_{36}$} & \colhead{Comment} \\
\colhead{}            & \colhead{(h m s J2000)} & \colhead{(\degree\ \arcmin\ \arcsec\ J2000)} &
\colhead{(mJy)}                                &
\colhead{}      & \colhead{}             &
\colhead{(mJy, uncorrected)}                             &
\colhead{(mJy, uncorrected)}                             &
\colhead{(mJy, uncorrected)}                            & \colhead{}
}
\startdata
SSEP70 J042854.2$-$555308 &  04 28 54.25   & $-$55 53 09.0 & $  55.6  \pm  1.7 $ &  34   &  1.3  & $ 37.0  \pm 1.1 $ & $ 57.3 \pm 1.8$ & $  70.6 \pm 2.3$ &   \\
SSEP70 J042820.1$-$555302 &  04 28 20.17   & $-$55 53 02.1 & $  57.3  \pm  1.4 $ &  42   &  0.30 & $ 36.14 \pm 0.88$ & $ 50.6 \pm 1.5$ & $  60.6 \pm 2.0$ &   \\
SSEP70 J042846.9$-$555102 &  04 28 46.95   & $-$55 51 02.4 & $  33.1  \pm  1.5 $ &  21   &  0.40 & $ 21.61 \pm 0.99$ & $ 26.8 \pm 1.7$ & $  22.3 \pm 2.2$ &   \\
SSEP70 J043006.1$-$554910 &  04 30 06.10   & $-$55 49 10.1 & $  43.9  \pm  1.5 $ &  29   &  0.60 & $ 36.31 \pm 0.98$ & $ 59.2 \pm 1.7$ & $  70.6 \pm 2.1$ & D \\
SSEP70 J043006.2$-$554852 &  04 30 06.30   & $-$55 48 52.3 & $  37.8  \pm  1.4 $ &  25   &  0.60 & $ 33.25 \pm 0.93$ & $ 58.4 \pm 1.6$ & $  70.0 \pm 2.1$ & D \\
SSEP70 J043133.4$-$554429 &  04 31 33.43   & $-$55 44 29.4 & $  34.1  \pm  1.6 $ &  21   &  0.85 & $ 22.0  \pm 1.1 $ & $ 29.6 \pm 1.8$ & $  36.1 \pm 2.2$ &   \\
SSEP70 J043157.5$-$554305 &  04 31 57.58   & $-$55 43 05.8 & $  41.4  \pm  1.5 $ &  26   &  0.83 & $ 25.1  \pm 1.0 $ & $ 30.3 \pm 1.8$ & $  37.4 \pm 2.3$ &   \\
SSEP70 J043246.9$-$554044 &  04 32 46.93   & $-$55 40 44.5 & $ 105.3  \pm  1.7 $ &  66   &  2.0  & $ 72.1  \pm 1.1 $ & $102.4 \pm 1.7$ & $ 119.5 \pm 2.1$ & D \\
SSEP70 J043248.6$-$554039 &  04 32 48.67   & $-$55 40 39.7 & $  18.5  \pm  1.5 $ &  12   &  2.0  & $ 41.3  \pm 1.0 $ & $ 88.1 \pm 1.7$ & $ 104.7 \pm 2.2$ & D \\
SSEP70 J042812.6$-$554601 &  04 28 12.68   & $-$55 46 01.9 & $  27.5  \pm  1.3 $ &  20   &  0.70 & $ 18.91 \pm 0.89$ & $ 23.4 \pm 1.5$ & $  29.5 \pm 2.0$ &   \\
SSEP70 J042951.3$-$554404 &  04 29 51.39   & $-$55 44 04.1 & $  46.0  \pm  1.7 $ &  27   &  0.46 & $ 29.7  \pm 1.1 $ & $ 40.1 \pm 1.9$ & $  44.5 \pm 2.3$ &   \\
SSEP70 J042905.9$-$554343 &  04 29 05.95   & $-$55 43 43.5 & $  58.4  \pm  1.6 $ &  38   &  0.53 & $ 36.2  \pm 1.0 $ & $ 45.1 \pm 1.7$ & $  49.9 \pm 2.1$ &   \\
SSEP70 J043050.2$-$554131 &  04 30 50.29   & $-$55 41 31.4 & $  45.3  \pm  1.6 $ &  29   &  0.40 & $ 30.8  \pm 1.1 $ & $ 47.3 \pm 1.7$ & $  55.5 \pm 2.1$ &   \\
SSEP70 J043347.3$-$553710 &  04 33 47.34   & $-$55 37 10.2 & $  26.2  \pm  1.5 $ &  16   &  0.30 & $ 18.7  \pm 1.0 $ & $ 29.9 \pm 1.8$ & $  43.4 \pm 2.3$ &   \\
SSEP70 J043017.0$-$554111 &  04 30 17.05   & $-$55 41 11.9 & $  33.4  \pm  1.4 $ &  23   &  0.63 & $ 21.95 \pm 0.97$ & $ 31.3 \pm 1.6$ & $  35.8 \pm 2.1$ &   \\
\enddata
\tablecomments{Table~\ref{tab:pcat70} is published in its entirety in the electronic edition of the {\it Astrophysical Journal Supplement Series}.  The first 15 entries are shown here for guidance regarding its form and content. The first column gives the source name in IAU format. The second and third columns list the RA and Dec for each source. The fourth column gives the PRF-fitted flux density and its formal uncertainty. The fifth column gives the SNR estimate, and the sixth column gives the reduced $\chi^2$ for the fit. Columns 7, 8, and 9 list the (uncorrected) aperture flux densities and uncertainties (\S\ref{ssec:apphot}) using 16\arcsec, 28\arcsec, and 36\arcsec\ radius apertures, respectively. The last column indicates sources that were passively deblended.}
\end{deluxetable}

\clearpage

\subsection{Point Response Function}
\label{ssec:prf}

\begin{figure*}
\begin{center}
\includegraphics[width=6.5in]{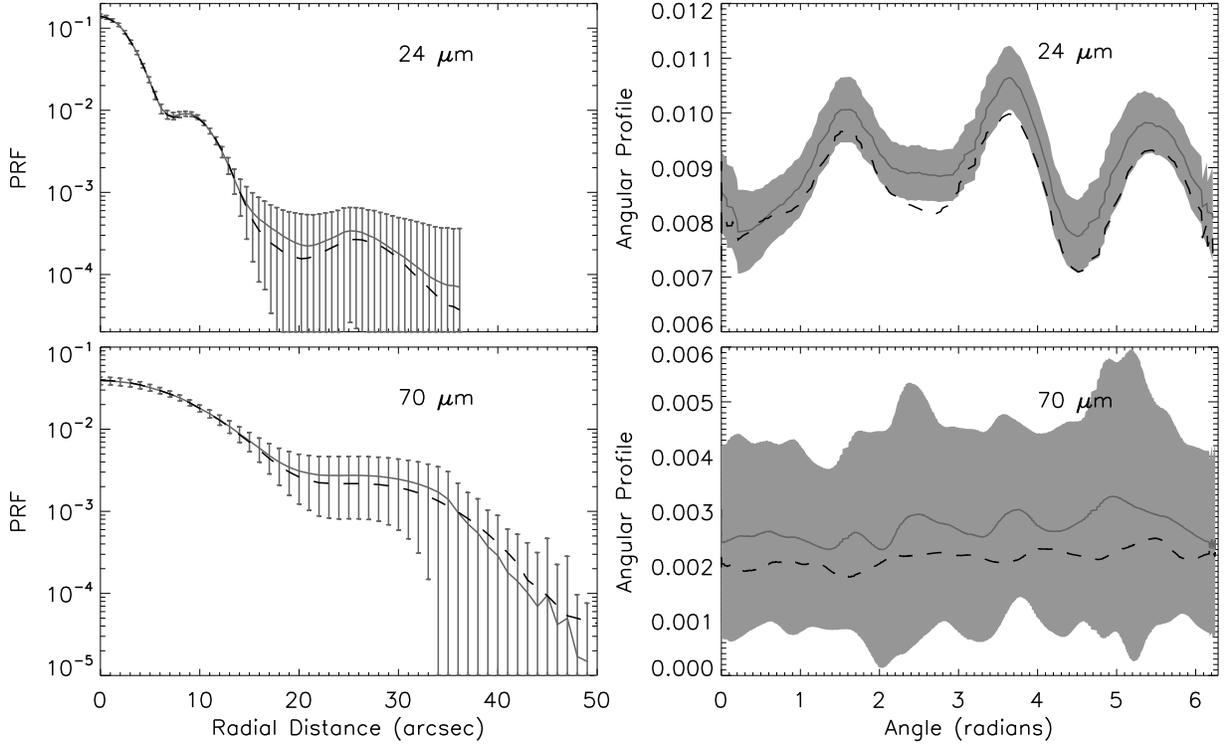}
\caption{Point response functions (PRFs) derived from the SEP {\it Spitzer} data, as described in \S\ref{ssec:prf}. The left panels show the radially averaged PRFs at 24\,\mum\ (top) and 70\,\mum\ (bottom). The gray solid curves are the PRFs estimated from our data using the {\sc APEX} PRF Estimate pipeline, where the error bars represent the standard deviation on these measurements. The dashed black curves are the radially averaged PRFs provided by the SSC. The right panels show the angular profiles of the PRFs at the first Airy ring. The solid gray curves and gray shaded regions are the PRF estimates from the SEP data and the standard deviation on these measurements, respectively, while the dashed black curves are from the PRFs provided by the SSC.}
\label{fig:prf} 
\end{center}
\end{figure*}

In fitting the source candidates to estimate their flux densities and positions, we use the 24 and 70\,\mum\ PRFs available from the SSC that had been produced using the {\sc MOPEX} PRF Estimate pipeline on previous data-sets.\footnote{http://ssc.spitzer.caltech.edu/mips/calibrationfiles/prfs/} We compare these to PRFs derived from our own data. For the 24\,\mum\ data, we use the PRF Estimate pipeline to cut and co-add postage-stamp images centered on 47 of the bright stars within this field, excluding those that are close to and/or confused with other bright sources, so as to get a clean estimate of the PRF. For the 70\,\mum\ data, we estimate the PRF by co-adding postage-stamp images centered on 129 70\,\mum\ sources detected with SNR $\ge50$ that are located far from other sources in the map and are not resolved galaxies. The radially averaged PRFs estimated from the SEP {\it Spitzer} data, and their angular profiles at the first Airy ring, are shown in Figure~\ref{fig:prf}. We find that the PRFs estimated from our data are in good agreement (within the measurement uncertainties) with the PRFs provided by the SSC, which are shown in Figure~\ref{fig:prf} for comparison. Since the latter are less noisy, we elect to use the PRF estimates from the SSC for point-source extraction and for all other analyses described below.

\subsection{Calibration and Color Corrections}
\label{ssec:cal}

The uncertainty in the absolute flux calibration of point sources is $\sim$4\% and $\sim$7\% at 24 and 70\,\mum, respectively \citep{engelbracht07,gordon07}. We use the default flux conversion factors from instrument units of 0.0447\,MJy\,sr$^{-1}$ at 24\,\mum\ and 702\,MJy\,sr$^{-1}$ at 70\,\mum, which are determined from frequent observations of primary and secondary calibrator stars assuming a blackbody spectrum with $T = 10\,000$\,K. Since extragalactic sources may have a very different spectrum across the 24 and 70\,\mum\ bands, color corrections should be applied to the flux densities listed in Tables~\ref{tab:pcat24}, \ref{tab:ecat}, and \ref{tab:pcat70}. However, given the range in spectral energy distributions (SEDs) and redshifts expected for different types of sources, it is difficult to choose a single template that will be appropriate for all sources. For this reason, we choose not to apply color corrections to the flux density measurements, and advise users of this catalog to compute their own (or alternatively, use the color corrections listed in the MIPS Instrument Handbook for an appropriate source spectrum) when needed.

\subsection{False Detection Rate}
\label{ssec:fdr}

The SNR derived from the data uncertainties does not represent the formal statistical significance of a detection under the assumption of Gaussian distributed noise. Therefore, we estimate the expected fraction of sources in our point-source catalogs that are false detections (i.e.~positive noise peaks) through simulation by running the same source-extraction algorithms described in \S\ref{ssec:cat24} and \S\ref{ssec:cat70} on our jackknifed noise realizations for the 24\,\mum\ (\S\ref{ssec:map24}) and 70\,\mum\ (\S\ref{ssec:map70}) maps. For the 24\,\mum\ catalog, we expect 1.8\% ($\sim1700$) of the sources listed in Table~\ref{tab:pcat24} to be false detections. For the 70\,\mum\ catalog, we expect 1.1\% ($\sim$9 to 10) of the sources listed in Table~\ref{tab:pcat70} to be false detections. Note, however, that due to the small number of sources detected in the 70\,\mum\ jackknife map
, this estimate is crude. Furthermore, for pixels with low coverage (i.e. where there is a small number of BCDs available for averaging), our jackknifing technique is less effective at removing the contribution from real sources. This can leave more pixels with excess positive or negative outliers than would be expected from pure noise, as can be seen by comparing the pixel flux distributions from the jackknifed maps to the simulated noise maps in Figures~\ref{fig:hist24} and \ref{fig:hist70}. For this reason, the false detection rates reported here should be considered upper limits.

\subsection{Completeness}
\label{ssec:comp}

\begin{figure*}
\begin{center}
\includegraphics[width=6.5in]{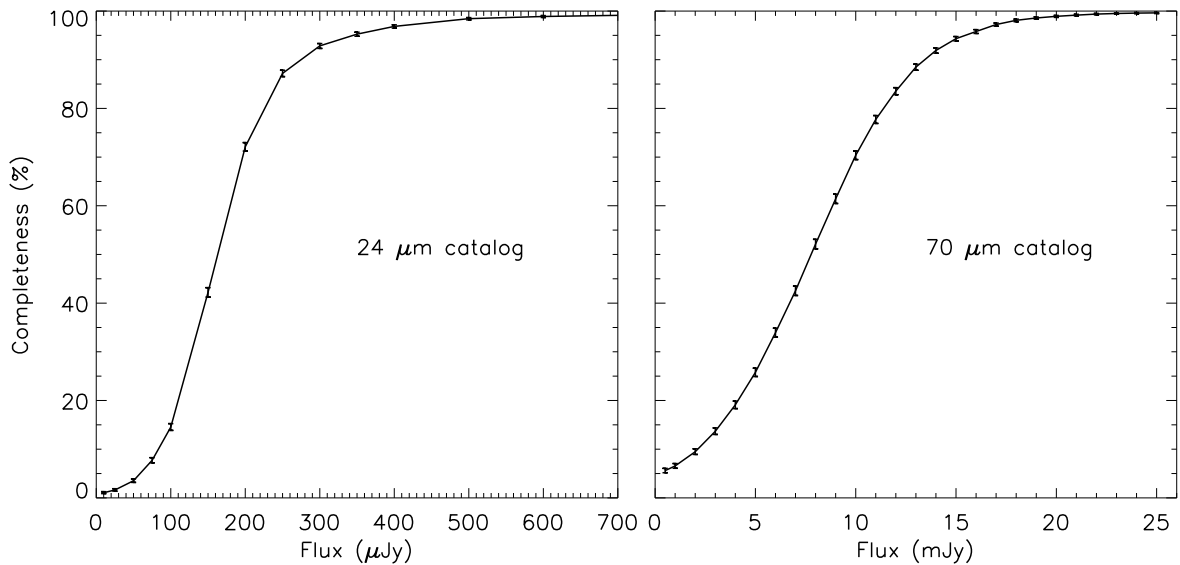}
\caption{The catalog completeness for the 24\,\mum\ (left) and 70\,\mum\ (right) catalogs. The error bars represent the 95\% confidence interval from the binomial distribution.}
\label{fig:comp} 
\end{center}
\end{figure*}

We estimate the 24\,\mum\ catalog completeness through simulation by injecting point sources with known flux density into the mosaic map and computing their recovery rate. We simultaneously inject 10\,000 simulated sources into the 24\,\mum\ mosaic map at discrete flux density values ranging from 10\,\muJy\ to 1200\,\muJy. Since the number density of simulated sources inserted at the same time is low, they do not appreciably change the noise properties of the map. At the same time, by inserting simulated sources into the real map, we account for the effects of confusion noise on the catalog completeness. To avoid contamination from the blending of $\ge2$ sources, every simulated source is injected $>7.7$\arcsec\ ($\sim2.5$ times the half-width at half maximum, HWHM) from any real source and from any other simulated source. We run the same source-extraction algorithm described in \S\ref{ssec:cat24} on these simulated maps; if an input source is detected with SNR $\ge5$ within 7.4\arcsec\ of its input position, it is considered to be recovered. The 24\,\mum\ catalog completeness as a function of intrinsic flux density is shown in Figure~\ref{fig:comp}. The catalog is 80\% complete at 230\,\muJy, and 95\% complete at 350\,\muJy.

We estimate the 70\,\mum\ catalog completeness in the same manner. For flux densities ranging from 0.5\,mJy to 25\,mJy, we simultaneously inject 10\,000 simulated sources into the mosaic map. Every simulated source is injected $>23$\arcsec\ ($\sim2.5$ times the HWHM) from any real 70\,\mum\ source and from other simulated sources. We run the source-extraction algorithm described in \S\ref{ssec:cat70} on these simulated maps, and consider an input source recovered if it is detected with SNR $\ge6$ within 20\arcsec\ of its input position. The 70\,\mum\ catalog completeness is shown in Figure~\ref{fig:comp}. The catalog is 80\% complete at 11\,mJy, and 95\% complete at 15\,mJy.

\subsection{Aperture Photometry}
\label{ssec:apphot}

We use {\sc APEX} to perform aperture photometry on the sources in this field in order to determine more accurate flux measurements for sources that are not well fit by the PRF and to aid in identifying extended sources. For the 24\,\mum\ sources, we use three different circular apertures with radii of 4.9\arcsec, 7.4\arcsec, and 15\arcsec\ (1.7, 2.5, and 5.0 times the HWHM of the 24\,\mum\ beam, respectively). For each source we estimate the background by computing the mode within an annulus of $20-32$\arcsec\ surrounding the source, and we subtract this background from the aperture fluxes. The aperture photometry for the 24\,\mum\ point sources is listed in columns $7-9$ of Table~\ref{tab:pcat24}. For the 70\,\mum\ point sources, we use three circular apertures with radii of 16\arcsec, 28\arcsec, and 36\arcsec\ (1.8, 3.1, and 4.0 times the HWHM of the 70\,\mum\ beam, respectively). We estimate and subtract the background, computed as the mode within an annulus of $60-80$\arcsec\ surrounding the source. The aperture photometry for each 70\,\mum\ point source is listed in columns $7-9$ of Table~\ref{tab:pcat70}. The uncertainties on the aperture fluxes represent the quadrature sum of the data uncertainties over the aperture area. For blank entries, the measured aperture fluxes were $<0$\,\muJy; in both the 24 and 70\,\mum\ cases, this occurs predominately for the largest radius aperture, while the smallest radius aperture always results in a net positive flux value. The aperture corrections are determined by integrating the PRFs, and are listed in Table~\ref{tab:apcorr} for easy reference.

\begin{deluxetable}{lcc}
\tabletypesize{\scriptsize}
\tablecaption{Aperture Corrections\label{tab:apcorr}}
\tablewidth{0pt}
\tablehead{
\colhead{} & \colhead{Aperture Radius} & \colhead{Aperture Correction} \\
\colhead{} & \colhead{\arcsec}         & \colhead{}
}
\startdata
24\,\mum &  4.9 & $1.84$ \\
         &  7.4 & $1.51$ \\
         & 15   & $1.08$ \\
70\,\mum & 16   & $1.57$ \\
         & 28   & $1.20$ \\
         & 36   & $1.05$ \\
\enddata
\tablecomments{To correct the measured aperture fluxes listed in Tables~\ref{tab:pcat24} and \ref{tab:pcat70}, multiply by the values in this table.}
\end{deluxetable}

Given that some regions of this field are crowded (mostly at 24\,\mum), which can affect both the aperture and background measurements, we recommend using the PRF-fitted flux densities for point sources that are well fit by the PRF. In other cases, it is generally a good idea to visually inspect the region surrounding the source of interest to decide which aperture is best to use, or to recalculate the flux using a different aperture and background annulus if needed.

\subsection{Extended Sources}
\label{ssec:ext}

A significant fraction of false detections come from extended sources being split into multiple pixel clusters during image segmentation. As an example, we show the 24 and 70\,\mum\ postage stamp images centered on the galaxy ESO 158-G 007 in the bottom panels of Figure~\ref{fig:fpexp}; the small circles and boxes indicate the ``sources'' initially identified by {\sc APEX}, where the boxes represent those arising from the extended emission of ESO 158-G 007. We visually inspect the 24 and 70\,\mum\ images and exclude such false detections from the point-source catalogs.

For a more rigorous analysis, we identify candidate extended sources by comparing the PRF-fitted fluxes to the aperture fluxes, using the 7.4\arcsec\ and 16\arcsec\ radius apertures for the 24 and 70\,\mum\ fluxes, respectively. If the PRF-fitted and (corrected) aperture fluxes do not agree within their $3\sigma$ uncertainties and the aperture flux is higher, the source is possibly extended. For sources that are well fit by the PRF ($\chi^2_\mathrm{r}\le3$), this is true for only 1.7\% (2.4\%) of the 24\,\mum\ (70\,\mum) sources and largely arises from multiple sources lying within the aperture radius. For sources with $\chi^2_\mathrm{r}>3$, 34\% (32\%) of the 24\,\mum\ (70\,\mum) sources meet this criterion. Therefore we believe that this criterion will select most of the resolved galaxies in this field.

At 24\,\mum\ there are a total of 758 candidate extended sources with $\chi^2_\mathrm{r}>3$. We also consider an additional 604 point sources that are well fit by the PRF, but whose PRF-fitted and apertures fluxes differ by more than $3\sigma$; these sources are often found in proximity to each other and could potentially arise from extended emission from a single source. We cross-check the positions of these sources with the NASA Extragalactic Database (NED). For those with an extragalactic counterpart, we visually compare the optical/near-IR images and the 24\,\mum\ emission in order to select an appropriate aperture size for measuring the 24\,\mum\ surface brightness. If the emission is contained within a 15\arcsec\ aperture, we do not remeasure the aperture photometry since this information is already given in Column 9 of Table~\ref{tab:pcat24}. For the 63 extended sources that require apertures with radii $>15$\arcsec, we first use the {\sc APEX} QA pipeline to subtract the 93\,098 point sources from the 24\,\mum\ mosaic; we then carry out aperture photometry on this residual map for each extended source using the appropriate aperture sizes and source positions from NED. This is demonstrated for ESO 158-G 007 in the bottom panels of Figure~\ref{fig:fpexp}, where the black cross marks the source position from NED, and the black dashed circle indicates the aperture radius (60\arcsec) used. The resolved galaxy catalog is given in Table~\ref{tab:ecat}. According to the MIPS Instrument Handbook, the total uncertainty on the flux calibration for extended sources is $\sim$15\%.

We carry out an independent check for candidate extended sources with the 70\,\mum\ catalog using the same criterion. There are 16 sources for which the measured aperture flux is larger than PRF-fitted flux and discrepant by $>3\sigma$. To this list, we add an additional six sources that do not meet this criterion, but by eye are clearly extended. For sources with an extragalactic counterpart found in NED, we pick out 8 extended sources that require apertures larger than 36\arcsec\ (i.e.~the largest aperture radius used on the point-source catalog). As with the 24\,\mum\ data we carry out aperture photometry for these extended sources after subtracting the 891 70\,\mum\ point sources from the map. These measurements are listed in column 5 of Table~\ref{tab:ecat}. For the remaining extended 24\,\mum\ sources, we list in Table~\ref{tab:ecat} the 70\,\mum\ PRF-fitted fluxes and uncertainties from the point-source catalog where available (Table~\ref{tab:pcat70}), and we note the 70\,\mum\ source identification in the last column.


\section{Conclusions}
\label{sec:conc}

We have imaged an 11.5\,deg$^2$ field towards the SEP at 24 and 70\,\mum\ with MIPS, achieving $1\sigma$ depths of $31-110$\,\muJy\,beam$^{-1}$ at 24\,\mum\ and $2.2-40$\,mJy\,beam$^{-1}$ at 70\,\mum. We identify 93\,098 point sources with SNR $\ge5$ at 24\,\mum, and 63 resolved galaxies. Through simulations, we determine that the 24\,\mum\ point-source catalog has an expected false detection rate of 1.8\%, and is 80\% complete at 230\,\muJy. From the 70\,\mum\ map, we identify 891 point sources with SNR $\ge6$; this 70\,\mum\ catalog is 80\% complete at 11\,mJy, with a false detection rate of 1.1\%.

We have made the 24 and 70\,\mum\ mosaic images, their corresponding uncertainty and coverage maps, and the catalogs described in this paper available to the public through the NASA/IPAC Infrared Science Archive (IRSA)\footnote{http://irsa.ipac.caltech.edu/} as a {\it Spitzer} contributed data-set, and through the BLAST public website\footnote{http://blastexperiment.info/release/SEP\_MIPS/sep-mips.php}.



\acknowledgments

We thank the anonymous referee for his/her suggestions, which improved the clarity of this paper. This work is based on observations made with the {\it Spitzer} Space Telescope, which is operated by the Jet Propulsion Laboratory, California Institute of Technology under a contract with NASA. Support for this work was provided by NASA through an award issued by JPL/Caltech. This research was supported by the Natural Sciences and Engineering Research Council of Canada and by the Canadian Space Agency. This research has made use of the NASA/IPAC Infrared Science Archive (IRSA) and the NASA/IPAC Extragalactic Database (NED), both of which are operated by the Jet Propulsion Laboratory, California Institute of Technology, under contract with the National Aeronautics and Space Administration. 



{\it Facilities:} \facility{{\it Spitzer} (MIPS)}

\bibliographystyle{apj}
\bibliography{SEP_MIPS24-70um_references}

\begin{thebibliography}{41}
\expandafter\ifx\csname natexlab\endcsname\relax\def\natexlab#1{#1}\fi

\bibitem[{{Aretxaga} {et~al.}(2007)}]{aretxaga07}
{Aretxaga}, I., {et~al.} 2007, \mnras, 379, 1571

\bibitem[{{Ashby} {et~al.}(2006)}]{ashby06}
{Ashby}, M.~L.~N., {et~al.} 2006, \apj, 644, 778

\bibitem[{{Austermann} {et~al.}(2010)}]{austermann10}
{Austermann}, J.~E., {et~al.} 2010, \mnras, 401, 160

\bibitem[{{Bertoldi} {et~al.}(2007)}]{bertoldi07}
{Bertoldi}, F., {et~al.} 2007, \apjs, 172, 132

\bibitem[{{Blain} {et~al.}(2002){Blain}, {Smail}, {Ivison}, {Kneib}, \&
  {Frayer}}]{blain02}
{Blain}, A.~W., {Smail}, I., {Ivison}, R.~J., {Kneib}, J., \& {Frayer}, D.~T.
  2002, \physrep, 369, 111

\bibitem[{{Borys} {et~al.}(2003){Borys}, {Chapman}, {Halpern}, \&
  {Scott}}]{borys03}
{Borys}, C., {Chapman}, S., {Halpern}, M., \& {Scott}, D. 2003, \mnras, 344,
  385

\bibitem[{{Chapin} {et~al.}(2009)}]{chapin09}
{Chapin}, E.~L., {et~al.} 2009, \mnras, 398, 1793

\bibitem[{{Chapin} {et~al.}(2010)}]{chapin10}
---. 2010, ArXiv e-prints

\bibitem[{{Chapman} {et~al.}(2005){Chapman}, {Blain}, {Smail}, \&
  {Ivison}}]{chapman05}
{Chapman}, S.~C., {Blain}, A.~W., {Smail}, I., \& {Ivison}, R.~J. 2005, \apj,
  622, 772

\bibitem[{{Coppin} {et~al.}(2006)}]{coppin06}
{Coppin}, K., {et~al.} 2006, \mnras, 372, 1621

\bibitem[{{Devlin} {et~al.}(2009)}]{devlin09}
{Devlin}, M.~J., {et~al.} 2009, \nat, 458, 737

\bibitem[{{Dye} {et~al.}(2008)}]{dye08}
{Dye}, S., {et~al.} 2008, \mnras, 386, 1107

\bibitem[{{Dye} {et~al.}(2009)}]{dye09}
---. 2009, \apj, 703, 285

\bibitem[{{Engelbracht} {et~al.}(2007)}]{engelbracht07}
{Engelbracht}, C.~W., {et~al.} 2007, \pasp, 119, 994

\bibitem[{{Fixsen} {et~al.}(1998){Fixsen}, {Dwek}, {Mather}, {Bennett}, \&
  {Shafer}}]{fixsen98}
{Fixsen}, D.~J., {Dwek}, E., {Mather}, J.~C., {Bennett}, C.~L., \& {Shafer},
  R.~A. 1998, \apj, 508, 123

\bibitem[{{Frayer} {et~al.}(2006{\natexlab{a}})}]{frayer06a}
{Frayer}, D.~T., {et~al.} 2006{\natexlab{a}}, \aj, 131, 250

\bibitem[{{Frayer} {et~al.}(2006{\natexlab{b}})}]{frayer06b}
---. 2006{\natexlab{b}}, \apjl, 647, L9

\bibitem[{{Frayer} {et~al.}(2009)}]{frayer09}
---. 2009, \aj, 138, 1261

\bibitem[{{Gordon} {et~al.}(2005)}]{gordon05}
{Gordon}, K.~D., {et~al.} 2005, \pasp, 117, 503

\bibitem[{{Gordon} {et~al.}(2007)}]{gordon07}
---. 2007, \pasp, 119, 1019

\bibitem[{{Greve} {et~al.}(2004){Greve}, {Ivison}, {Bertoldi}, {Stevens},
  {Dunlop}, {Lutz}, \& {Carilli}}]{greve04}
{Greve}, T.~R., {Ivison}, R.~J., {Bertoldi}, F., {Stevens}, J.~A., {Dunlop},
  J.~S., {Lutz}, D., \& {Carilli}, C.~L. 2004, \mnras, 354, 779

\bibitem[{{Greve} {et~al.}(2008){Greve}, {Pope}, {Scott}, {Ivison}, {Borys},
  {Conselice}, \& {Bertoldi}}]{greve08}
{Greve}, T.~R., {Pope}, A., {Scott}, D., {Ivison}, R.~J., {Borys}, C.,
  {Conselice}, C.~J., \& {Bertoldi}, F. 2008, \mnras, 389, 1489

\bibitem[{{Hainline} {et~al.}(2009){Hainline}, {Blain}, {Smail}, {Frayer},
  {Chapman}, {Ivison}, \& {Alexander}}]{hainline09}
{Hainline}, L.~J., {Blain}, A.~W., {Smail}, I., {Frayer}, D.~T., {Chapman},
  S.~C., {Ivison}, R.~J., \& {Alexander}, D.~M. 2009, \apj, 699, 1610

\bibitem[{{Hauser} {et~al.}(1998)}]{hauser98}
{Hauser}, M.~G., {et~al.} 1998, \apj, 508, 25

\bibitem[{{Laurent} {et~al.}(2005)}]{laurent05}
{Laurent}, G.~T., {et~al.} 2005, \apj, 623, 742

\bibitem[{{Marsden} {et~al.}(2009)}]{marsden09}
{Marsden}, G., {et~al.} 2009, \apj, 707, 1729

\bibitem[{{Masci} {et~al.}(2005){Masci}, {Laher}, {Fang}, {Fowler}, {Lee},
  {Stolovy}, {Padgett}, \& {Moshir}}]{masci05}
{Masci}, F.~J., {Laher}, R., {Fang}, F., {Fowler}, J.~W., {Lee}, W., {Stolovy},
  S., {Padgett}, D., \& {Moshir}, M. 2005, in Astronomical Society of the
  Pacific Conference Series, Vol. 347, Astronomical Data Analysis Software and
  Systems XIV, ed. {P.~Shopbell, M.~Britton, \& R.~Ebert}, 468--+

\bibitem[{{Matsuhara} {et~al.}(2006)}]{matsuhara06}
{Matsuhara}, H., {et~al.} 2006, \pasj, 58, 673

\bibitem[{{Papovich} {et~al.}(2004)}]{papovich04}
{Papovich}, C., {et~al.} 2004, \apjs, 154, 70

\bibitem[{{Pascale} {et~al.}(2008)}]{pascale08}
{Pascale}, E., {et~al.} 2008, \apj, 681, 400

\bibitem[{{Pascale} {et~al.}(2009)}]{pascale09}
---. 2009, \apj, 707, 1740

\bibitem[{{Perera} {et~al.}(2008)}]{perera08}
{Perera}, T.~A., {et~al.} 2008, \mnras, 391, 1227

\bibitem[{{Pope} {et~al.}(2006)}]{pope06}
{Pope}, A., {et~al.} 2006, \mnras, 370, 1185

\bibitem[{{Puget} {et~al.}(1996){Puget}, {Abergel}, {Bernard}, {Boulanger},
  {Burton}, {Desert}, \& {Hartmann}}]{puget96}
{Puget}, J., {Abergel}, A., {Bernard}, J., {Boulanger}, F., {Burton}, W.~B.,
  {Desert}, F., \& {Hartmann}, D. 1996, \aap, 308, L5+

\bibitem[{{Rieke} {et~al.}(2004)}]{rieke2004}
{Rieke}, G.~H., {et~al.} 2004, \apjs, 154, 25

\bibitem[{{Sanders} {et~al.}(2007)}]{sanders07}
{Sanders}, D.~B., {et~al.} 2007, \apjs, 172, 86

\bibitem[{{Scott} {et~al.}(2008)}]{scott08}
{Scott}, K.~S., {et~al.} 2008, \mnras, 385, 2225

\bibitem[{{Scott} {et~al.}(2010)}]{scott10}
---. 2010, \mnras, 684

\bibitem[{{Scott} {et~al.}(2002)}]{scott02}
{Scott}, S.~E., {et~al.} 2002, \mnras, 331, 817

\bibitem[{{Wang} {et~al.}(2006){Wang}, {Cowie}, \& {Barger}}]{wang06}
{Wang}, W.-H., {Cowie}, L.~L., \& {Barger}, A.~J. 2006, \apj, 647, 74

\bibitem[{{Wei{\ss}} {et~al.}(2009)}]{weiss09}
{Wei{\ss}}, A., {et~al.} 2009, \apj, 707, 1201

\end{thebibliography}

\end{document}